\documentclass[prd,aps,twocolumn,showpacs,floats,floatfix]{revtex4}
\usepackage[dvips]{graphicx}
\usepackage{amsmath}
\usepackage{subfigure}
\usepackage{appendix}

\begin{document}

\title{Dark-matter admixed white dwarfs}

\author{S.-C. Leung, M.-C. Chu, L.-M. Lin, K.-W. Wong}
\affiliation{Department of Physics and Institute
of Theoretical Physics, The Chinese University
of Hong Kong, Hong Kong, China}

\date{\today}


\begin{abstract}

We study the equilibrium structures of white dwarfs with dark matter cores
formed by non-self-annihilating dark matter (DM) particles with mass ranging 
from 1 GeV to 100 GeV, which are assumed to form an ideal degenerate 
Fermi gas inside the stars.
For DM particles of mass 10 GeV and 100 GeV, we find that stable stellar models
exist only if the mass of the DM core inside the star is less than 
$O(10^{-3}) \ M_\odot$ and $O(10^{-6}) \ M_\odot$, respectively. The global
properties of these stars, and in particular the corresponding Chandrasekhar
mass limits, are essentially the same as those of traditional white dwarf
models without DM. Nevertheless, in the 10 GeV case, the gravitational 
attraction of the DM core is strong enough to squeeze the normal matter in 
the core region to densities above neutron drip, far above those in 
traditional white dwarfs. 
For DM with particle mass 1 GeV, the DM core inside the star
can be as massive as $\sim 0.1 M_\odot$ 
and affects the global structure of the star significantly.
In this case, the radius of a stellar model with DM can be 
about two times smaller than that of a traditional white dwarf. 
Furthermore, the Chandrasekhar mass limit can also be decreased by as much as 
40\%. Our results may have implications on to what extent type Ia supernovae 
can be regarded as standard candles - a key assumption in the discovery of 
dark energy.

\end{abstract}


\maketitle


\section{Introduction}
\label{sec:introduction}

It has been widely accepted that more
than 80\% of matter in the universe is dark
matter (DM), most of which believed to be
non-baryonic, possibly weakly interacting massive particles (WIMP). 
Compelling evidences for DM include observations of the flatness of 
galactic rotation curves, measurements of the Cosmic microwave 
background and baryon-acoustic oscillations
(see, e.g., \cite{Freese2008, DAmico2009, Roos2010} for reviews). 
Nevertheless, the properties of DM particles, including their spin, mass and 
interactions, are still largely unknown.

The effects of various DM candidates on stellar 
evolution and structure have been discussed in the literature. 
It is hoped that constraints on the properties of DM particles may be
obtained from observations of stellar objects such as the sun or compact 
stars. Such studies can be divided into two classes:
non-self-annihilating DM and self-annihilating DM.

Self-annihilating DM affects a star by supplying energy
through their annihilations into photons. The  
role of DM in the first generation stars as stellar seeds, together with 
the possibility of DM annihilation as the energy source in the first phase of
stellar evolution are discussed 
in \cite{Spolyar2008, Spolyar2009, Gondolo2010, Ripamonti2010, Hirano2011}.
The DM annihilation energy might provide 
sufficient pressure to delay
the star from entering the main-sequence stage.
Also, the main-sequence lifetime can be extended 
\cite{Casanellas2009}. It was pointed out in  
\cite{Fairbairn2008, Casanellas2011} that the
DM self-annihilation energy may broaden the path in 
the {\rm H-R} diagram. By considering DM self-annihilation 
in the sun, authors of \cite{Cumberbatch2010, Taoso2010}
suggest that solar neutrino flux and helioseismology
data can be used to constrain DM particle properties.
In compact stars, DM annihilation becomes the 
only heat-generation mechanism,
and their cooling curves or luminosity 
may be altered \cite{Moskalenko2007, 
Kouvaris2008, Lavallaz2010, Kouvaris:2, 
Hooper2010, McCullough2010, Fan2011}, which may in turn 
provide information on the DM properties,
scattering cross-sections, as well as 
local DM distributions. On the other hand, it has also been 
suggested that self-annihilating neutralino DM stars cannot exist \cite{Dai2009}.

Non-self-annihilating DM affects a star through 
its gravity by accumulating in the stellar core, or 
through its cooling of stellar materials by 
scattering with ordinary particles.  
In \cite{Zentner2011, Iocco2012} it is suggested 
that the cooling by DM scattering may increase 
the minimum mass for hydrogen-burning stars and a longer
main-sequence lifetime. It has also been suggested 
\cite{Frandsen2010} that the solar composition problem 
may be solved by admixing non-self-annihilating DM.
The changes in the orbits of stellar objects 
due to the increase of stellar mass 
from accreted DM are studied in
\cite{Peter2009, Iorio2010a, Iorio2010b}.
Convection zone profile may also be changed, 
bringing impact on the helium flash \cite{Dearborn1990}. 
Similar studies are applied to compact stars. 
The first study is pioneered in 1989 by 
Goldman and Nussinov \cite{Goldman1989},
which sets a limit on DM particle mass and
scattering cross-section by considering
very old compact stars. The idea of using 
compact stars to probe the DM 
particle properties is further elaborated in 
\cite{Bertone:1, Lavallaz2010, McCullough2010, Fan2011}.
Limits on the mass and scattering cross-sections
for different classes of DM, such as 
bosonic and fermionic DM with different scattering 
channels, are derived from compact star observations
\cite{Kouvaris2011a, Kouvaris2011b, Kouvaris2012, McDermott2012}.
In \cite{Kaplan:1, Sandin2009, Ciarcelluti2011},
models of non-self-annihilating DM, 
including asymmetric dark matter and mirror matter,
are examined. It is found that
a more compact neutron star is resulted when a
DM core is included. The gravity from the accumulated DM
might even be strong enough to trigger a phase transition
from nuclear matter to quark matter inside the star
and produce a gamma-ray burst \cite{Stone2010}.





In our previous work \cite{DANS1, DANS2}, we considered non-self-annihilating 
DM particles of mass $\sim 1$ GeV to study the equilibrium structure and 
radial oscillations of DM admixed neutron stars using a general relativistic
two-fluid formulation. In particular, we found a new class of compact stars 
which consists of a small normal matter (NM) core with radius of a few kilometers embedded in 
a ten-kilometer sized DM halo. Here NM refers to ordinary particles in the Standard Model.

In this paper, we extend our investigation by employing the two-fluid 
formulation to study white dwarfs (WD) with DM cores. 
We shall in general refer to these stellar models as hybrid white dwarfs (HWD) 
in the following. 
The impact of annihilating DM on the cooling of traditional WD 
is well studied \cite{Moskalenko2007, Bertone:1, McCullough2010, 
Hooper2010, Fan2011, Kouvaris:2, Yang2012}. 
However, the study of non-self-annihilating DM on the structure of WD is a 
relatively unexplored area. 
A recent study using bosonic condensate DM can be found in \cite{Li2012}.  
If the mass and/or radius of traditional WD near the Chandrasekhar limit 
are altered significantly due to the presence of DM cores,
the initial conditions of Type Ia supernova explosions would be affected, 
making it doubtful whether they can still be regarded as standard candles - a 
key assumption in the discovery of dark energy.

In this work, we shall study the equilibrium structure of HWD by 
assuming that the DM particles are non-self-annihilating fermionic 
particles with particle mass $m_{DM}$ ranging from 1 to 100 GeV.
The mass range is chosen based on two reasons. First, the most popular DM 
candidate, WIMP, is expected to have a mass of the order 100 GeV. {\it Second, 
data from the DAMA, CoGeNT, and CRESST experiments \cite{DAMA2008, CoGeNT2011, CRESST2012} 
are consistent with detecting light DM particles with a few GeV, though the results 
are in conflict with the null results reported by CDMS and XENON \cite{CDMS2011, XENON2012}.
More recently, the CDMS-II collaboration has also reported signals 
that are consistent with DM particles with mass $\sim 9$ GeV \cite{CDMS2013}.}

The major difference between our current work and our previous work 
\cite{DANS1, DANS2} lies in the NM equation
of state (EOS). The pressure inside a HWD is due mainly to a
degenerate electron gas instead of nuclear matter. 
Apart from the EOS, the typical 
length scales of NM and DM in a HWD differ
by many orders of magnitude. As we shall see below, for DM 
particles with mass 100 GeV, the typical size of a DM core
can be smaller than the radius of the whole star by a 
factor of $10^6$.
The outline of the paper is as follows: In Sec.~\ref{sec:model}, we briefly
outline the formulation for constructing HWD.
In Sec.~\ref{sec:results}, we study the structure of HWD for different DM
particle masses in detail.
Sec.~\ref{sec:discussion} summarizes our
results and discusses possible future investigation.
Finally, Appendix~\ref{sec:Appendix} discusses briefly the radial
oscillation modes of HWD. Appendix B discusses how our proposed
HWD might be formed. 
We use units where $G=c=1$ unless otherwise noted.


\section{Formulation}
\label{sec:model}

In our previous work \cite{DANS1,DANS2}, we study the structure and 
oscillations of DM admixed neutron stars using a general relativistic 
two-fluid formalism. 
Here we adopt the formulation to study HWD with DM cores.
The essential structure equations and numerical technique
for constructing a two-fluid star can be found in \cite{DANS2} (see also
\cite{Comer1999} for the full derivations). Here we only outline the essential
equations.

For a static and spherically symmetric spacetime $ds^2 = - e^{\nu(r)} dt^2 +
e^{\lambda(r)}dr^2 + r^2 ( d\theta^2 + \sin^2\theta d\phi^2)$, the structure
equations for a two-fluid compact star are given by \cite{Comer1999}
\begin{eqnarray}
A^0_0 p' + B^0_0 n' + \frac{1}{2} (B n + A p) \nu ' = 0,
\nonumber\\
C^0_0 p' + A^0_0 n' + \frac{1}{2} (A n + C p) \nu ' = 0,
\nonumber\\
\lambda^{'} = { {1 - e^\lambda} \over r } - 8\pi r e^\lambda \Lambda ,
\nonumber\\
\nu^{'} = - { {1 - e^\lambda} \over r } + 8\pi r e^\lambda \Psi ,
\label{eq:fluid_eq}
\end{eqnarray}
where $n$ and $p$ are the number densities of NM and DM, respectively.
The primes refer to derivatives with respect to $r$, and the coefficients
$A$, $B$, $C$, $A^0_0$, $B^0_0$, and $C^0_0$ are functions of the master
function $\Lambda$, which is the negative of the thermodynamics energy
density. The generalized pressure $\Psi$ is calculated from the master function
$\Lambda$. We refer the reader to \cite{Comer1999,DANS2} for the explicit 
expressions.

In the two-fluid formalism, the master function $\Lambda$ plays the role of
the EOS information needed in the structure calculation.
In this work, we assume that DM couples with NM only through
gravity. Hence, the master function is separable in the sense that
\begin{equation}
\Lambda(n,p) = \Lambda_{\rm NM}(n) + \Lambda_{\rm DM} (p) ,
\end{equation}
$\Lambda_{\rm NM} (n)$ and $\Lambda_{\rm DM} (p)$ being the negative of energy
densities of NM and DM, respectively.

To model the NM, we choose the Akmal-Pandharipande-Ravenhall 
EOS \cite{APR} to describe the high-density
nuclear matter. As we shall see in Sec.~\ref{sec:results}, we need to
model matter in the nuclear density range ($\sim 10^{14}\ {\rm g\ cm}^{-3}$)
because the NM in the core of a HWD can indeed reach this density range.
At lower densities we use the SLY4 \cite{SLY} and 
Baym-Pethick-Sutherland EOS \cite{BPS}.
On the other hand, we consider degenerate ideal Fermi gas EOS for DM.
We shall consider DM in the mass range from 1 GeV to 100 GeV.

\section{Results}
\label{sec:results}

In this section, we study the stellar properties of HWD with different
DM core masses $M_{{\rm DM}}$ and particle mass $m_{{\rm DM}}$.
Before presenting our results in detail, let us first give a brief
summary of our finding.

Since the DM core is described by an ideal Fermi gas, and it is well known
that the maximum stable mass of a self-gravitating Fermi gas depends on
the particle mass $m_{{\rm DM}}$, thus the maximum amount of DM that can exist
inside a stable HWD is determined by $m_{{\rm DM}}$.
As we shall see below, a stable HWD can only have a tiny DM core 
($M_{{\rm DM}} \sim 10^{-6}M_\odot$) if the core is composed of massive DM 
particles with $m_{DM} = 100$ GeV.
If the mass of the DM core increases beyond the maximum stable limit, the whole
star would collapse promptly to a black hole.
On the other hand, for low-mass DM particle $m_{{\rm DM}} = 1$ GeV, the
DM core can reach the level $M_{{\rm DM}} \sim 0.1 M_\odot$ and affect the 
structure of the star significantly. 
The relation between $m_{{\rm DM}}$
and the maximum DM core mass can be obtained from the energy
argument of Landau (see, e.g. \cite{Shapiro2008}), from 
which we have the scaling relation $M_{{\rm DM(max)}} \sim m_{DM}^{-2}$.

For $m_{{\rm DM}} > 100$ GeV, 
the DM core becomes so small in spatial size 
that it is difficult to be resolved.  
However, extrapolating our results to $m_{{\rm DM}} > 100$ GeV, 
we expect almost no change to the mass-radius relation.

\subsection{Chandrasekhar mass limit and Moon-sized HWD}

\begin{figure}
\centering
\includegraphics*[width=8cm,height=6cm]{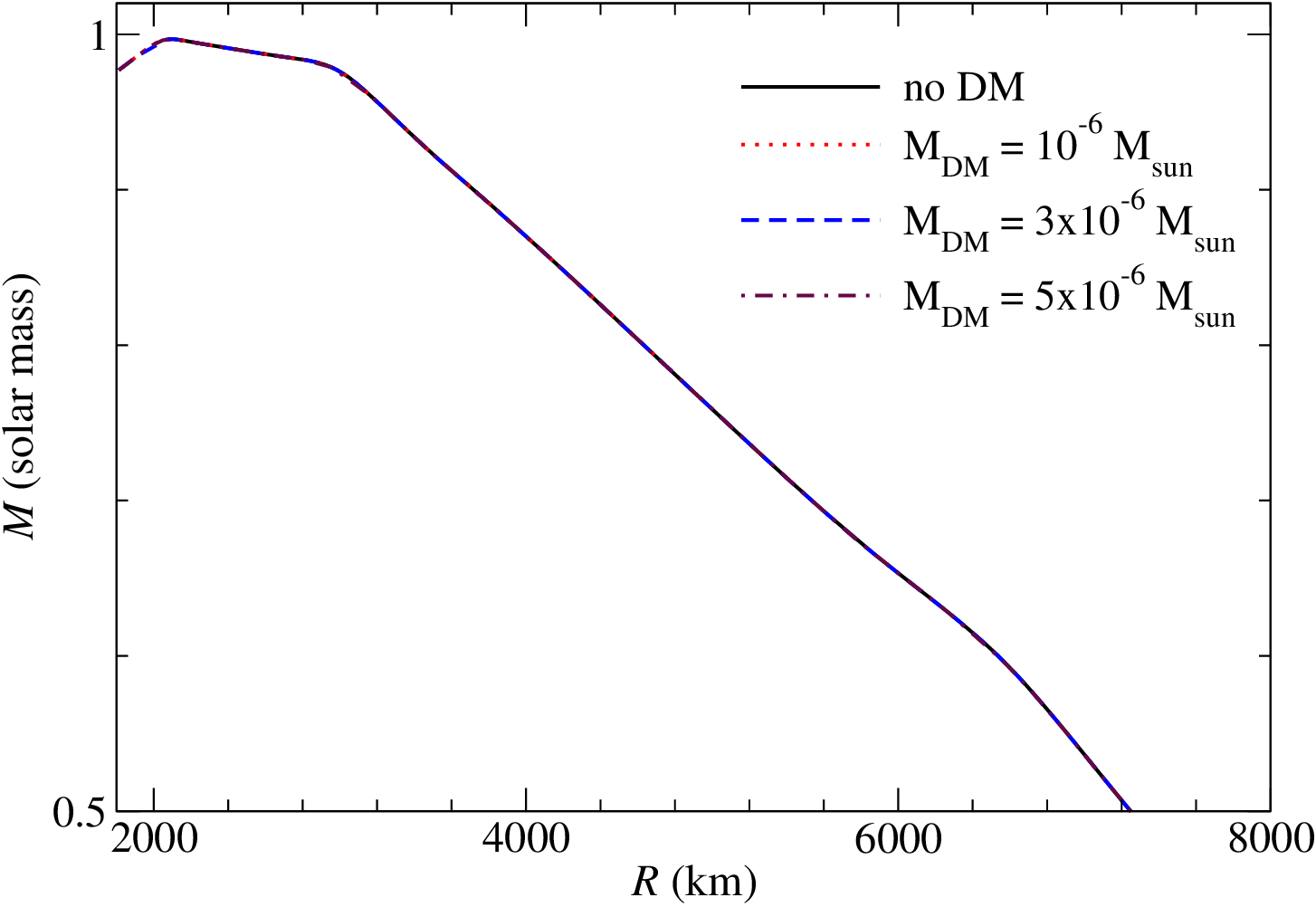}
\caption{Mass-radius relations of HWD for different DM core mass
$M_{DM}$. The DM particle mass is $m_{{\rm DM}}=100$ GeV.
We choose $M_{{\rm DM}}$ up to the value $5\times 10^{-6} M_\odot$, beyond which
no stable DM core can exist.  }
\label{fig:mr100}
\end{figure}

Here we study the mass-radius relation of HWD composed of DM with
different particle mass $m_{{\rm DM}}$ ranging from 1 GeV to 100 GeV.
In Fig.~\ref{fig:mr100}, we plot the mass-radius relations of HWD with DM core
formed by $m_{{\rm DM}}=100$ GeV DM particles. Results for three different
DM core mass $M_{DM}$ are plotted together with the case without DM.
For these massive DM particles, we find that HWD models cannot be constructed
with $M_{{\rm DM}} \gtrsim 5\times 10^{-6} M_\odot$, the value of which is set
by the maximum mass limit of a self-gravitating degenerate Fermi gas. It
should be noted that the mass of the NM fluid inside the DM core also
decreases the stability of the DM core. The total masses of these HWD models
are dominated by the NM fluid, and hence the DM cores have negligible effects
on the stellar structures as shown in Fig.~\ref{fig:mr100}.

Decreasing the DM particle mass can increase the maximum
stable mass limit of the degenerate DM cores inside HWD. One may thus expect
to see a significant difference between HWD and traditional WD models in the
low $m_{{\rm DM}}$ regime. 

\begin{figure}
\centering
\includegraphics*[width=8cm,height=6cm]{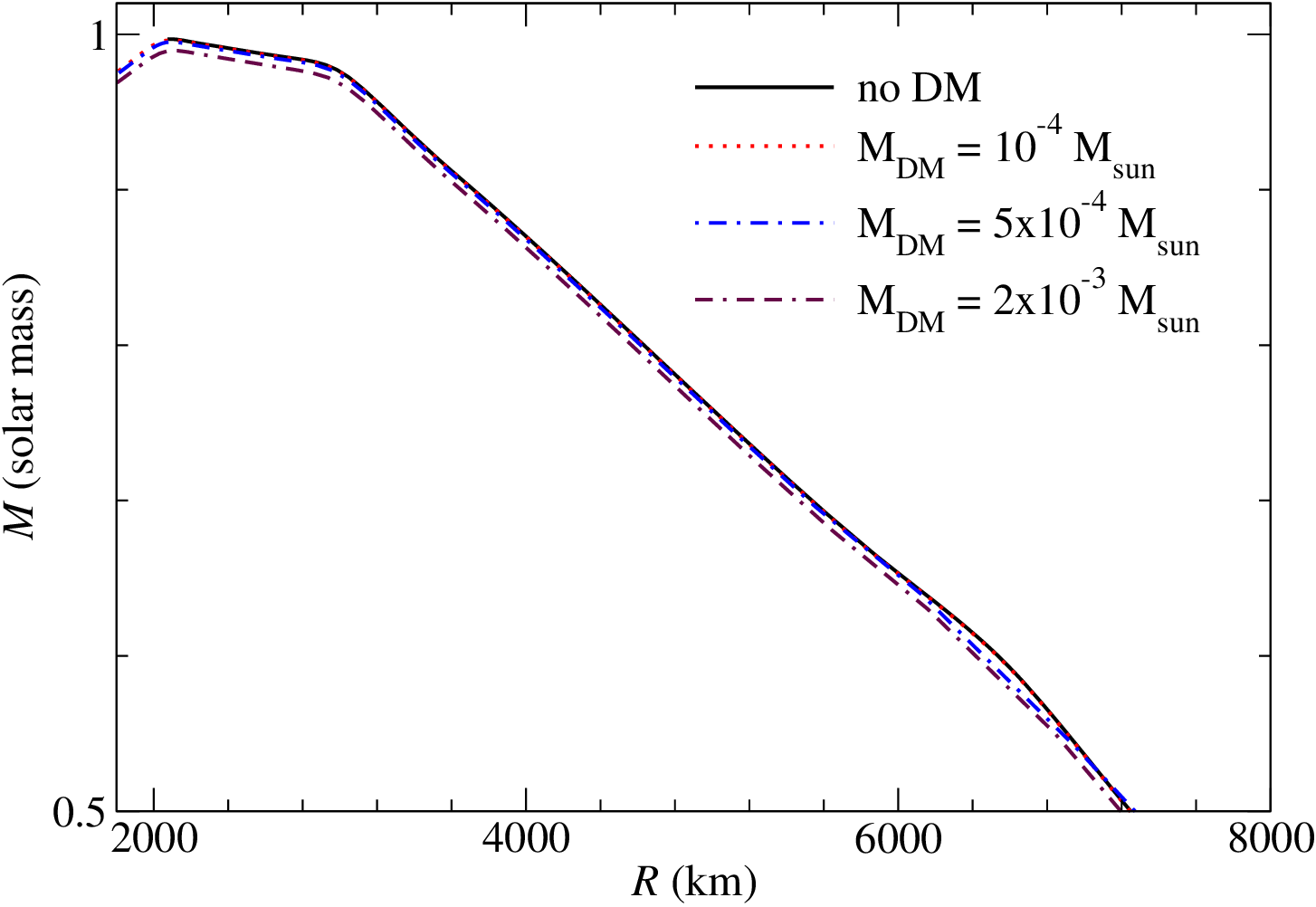}
\caption{Same as Fig. \ref{fig:mr100}, but for $m_{{\rm DM}} = 10$ GeV. }
\label{fig:mr10}
\end{figure}

In Fig.~\ref{fig:mr10}, we show the mass-radius relation of HWD for 
$m_{{\rm DM}}=10$ GeV.
We see that the maximum stable mass of the DM core increases to about
$M_{{\rm DM}} = 2 \times 10^{-3} M_\odot$, beyond which no HWD model can be
constructed. Although the DM core can now be more massive than that in the case
of $m_{{\rm DM}} = 100$ GeV, it is still not large enough to change the mass-radius
relation significantly. In particular, the DM core has little effect on the
Chandrasekhar mass limit of WD.
Hence, for non-self-annihilating massive DM particles, we 
conclude that stable HWD can only have tiny DM cores ($M_{{\rm DM}} << M$). 
As a result, the global properties of these stellar models are 
very similar to traditional WD.

\begin{figure}
\centering
\includegraphics*[width=8cm,height=6cm]{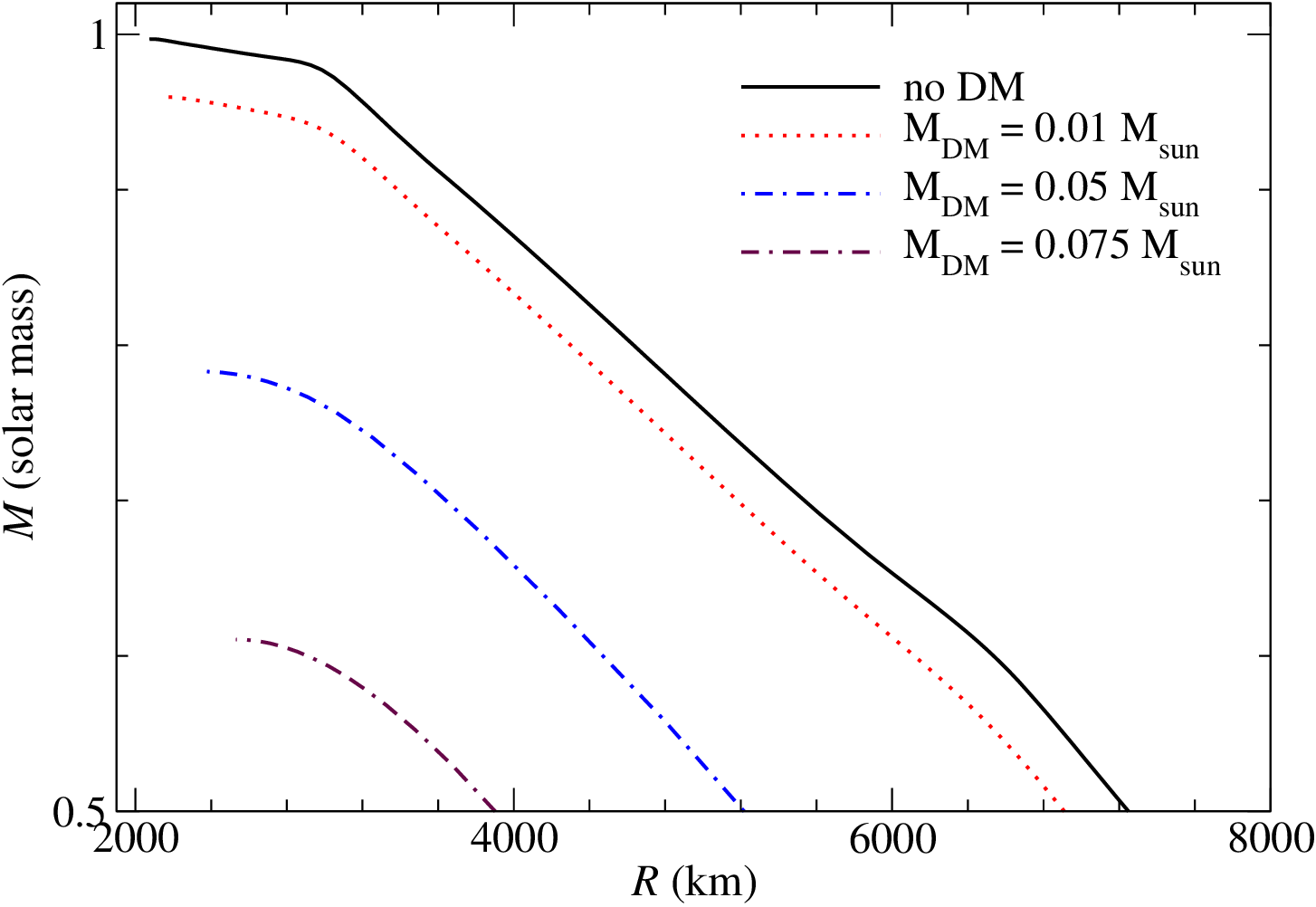}
\caption{Same as Fig.~\ref{fig:mr100}, but for $m_{{\rm DM}}=1$ GeV.}
\label{fig:mr1}
\end{figure}

Next we study the effects of low-mass DM with $m_{{\rm DM}}=1$ GeV. We plot
the corresponding mass-radius relation in Fig.~\ref{fig:mr1}. The maximum
stable mass limit of the DM core is $M_{{\rm DM}}=0.075 M_\odot$ in this case.
In contrast to the previous cases $m_{{\rm DM}}=10$ and 100 GeV, we now see
that the mass-radius relation of HWD depends sensitively on $M_{{\rm DM}}$.
In particular, the Chandrasekhar mass limit decreases from about
$1 M_\odot$ to $0.6 M_\odot$ as $M_{{\rm DM}}$ increases from 0 to $0.075 M_\odot$.
However, the radius of the stellar model corresponding to the
Chandrasekhar mass limit does not depend sensitively on $M_{{\rm DM}}$.
It should be noted that while both  
fluids contribute to the gravitational
potential, the pressure of each fluid does
not support the other one. Therefore, extra pressure gradient
is needed for one fluid to balance
the extra gravitational force exerted by the other fluid.
However, because the DM core is much more compact than the NM,
a slight increase in $M_{{\rm DM}}$ results in a large gravitational
attraction in the core and induces the collapse of the 
Chandrasekhar-mass model. In order to achieve
a new stable equilibrium model, the HWD should have a
much lower $M_{{\rm NM}}$. Note that the total mass of a HWD
is dominated by the NM, thus the Chandrasekhar mass
drops significantly with a slight increment in $M_{{\rm DM}}$.
On the contrary, the radius of the Chandrasekhar-mass model
displays a mild change. Along the sequence of Chandrasekhar-mass 
models, the increase in $M_{{\rm DM}}$ leads to a smaller radius because of
a stronger gravitational attraction. On the other hand, the
decrease in $M_{{\rm NM}}$ makes a HWD larger.
As these factors partially cancel each other, the radius of
the Chandrasekhar-mass model is thus insensitive to $M_{{\rm DM}}$.

One may also notice that, for a given total mass $M$, the radius
of the star decreases significantly as the amount of DM increases.
Fig.~\ref{fig:mr1} shows that the radius of a 0.6 $M_\odot$ traditional WD
without DM is about 6600 km. Due to the strong gravity of the DM core,
a HWD with the same total mass and $M_{{\rm DM}}=0.075 M_\odot$ has a radius of 
2800 km only.
For comparison, the radius of the moon is about 1700 km, whereas the
largest satellite in the solar system, Ganymede, has a radius of about 2600 km.

\subsection{HWD with central densities above neutron drip}

Since DM and NM are assumed to couple with each other only through gravity
in our calculations, the NM is supported by its own pressure against
gravitational attraction, which is sourced by both fluids.
Therefore, comparing to traditional WD models without DM, the presence of a
DM core in a HWD can provide extra gravitational attraction to squeeze the NM
in the core region to a higher density.

It is known that the global structure of a traditional WD is determined by 
the EOS below neutron drip
($\rho_{\rm drip}\approx 4\times 10^{11}\ {\rm g\ cm}^{-3}$), while that of
a neutron star is determined mainly by the EOS near or above nuclear density
($\rho_{\rm nuc}\approx 2.8\times 10^{14}\ {\rm g\ cm}^{-3}$).
In the domain between $\rho_{\rm drip}$ and $\rho_{\rm nuc}$, the nuclei
become more neutron rich and the electron gas degeneracy pressure drops
significantly as the density increases. As a result, no stable traditional WD
can have central densities in this range.
Here we show that, in the presence of a DM core, it is indeed possible for
the NM inside stable HWD to have a central density in this domain.

\begin{figure}
\centering
\includegraphics*[width=8cm,height=6cm]{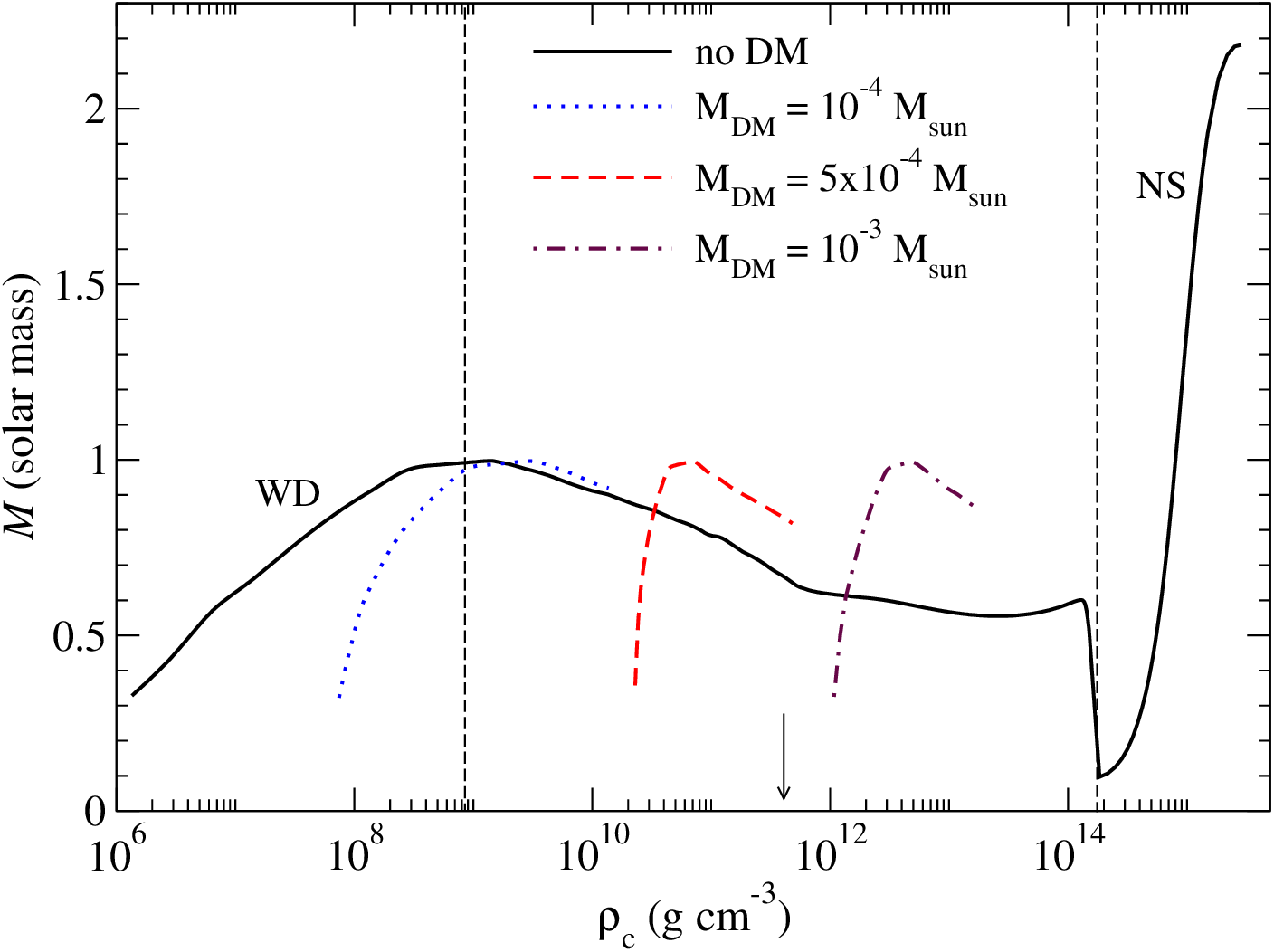}
\caption{Total mass $M$ is plotted against the NM central density
$\rho_c$ for $m_{{\rm DM}}=10$ GeV. 
The region between the two vertical dashed lines is the ``forbidden''
region in which no stable traditional white dwarf can exist. 
The arrow in the figure indicates the neutron drip density.}
\label{fig:Mrhoc_dm10GeV}
\end{figure}

\begin{figure}
\centering
\includegraphics*[width=8cm,height=6cm]{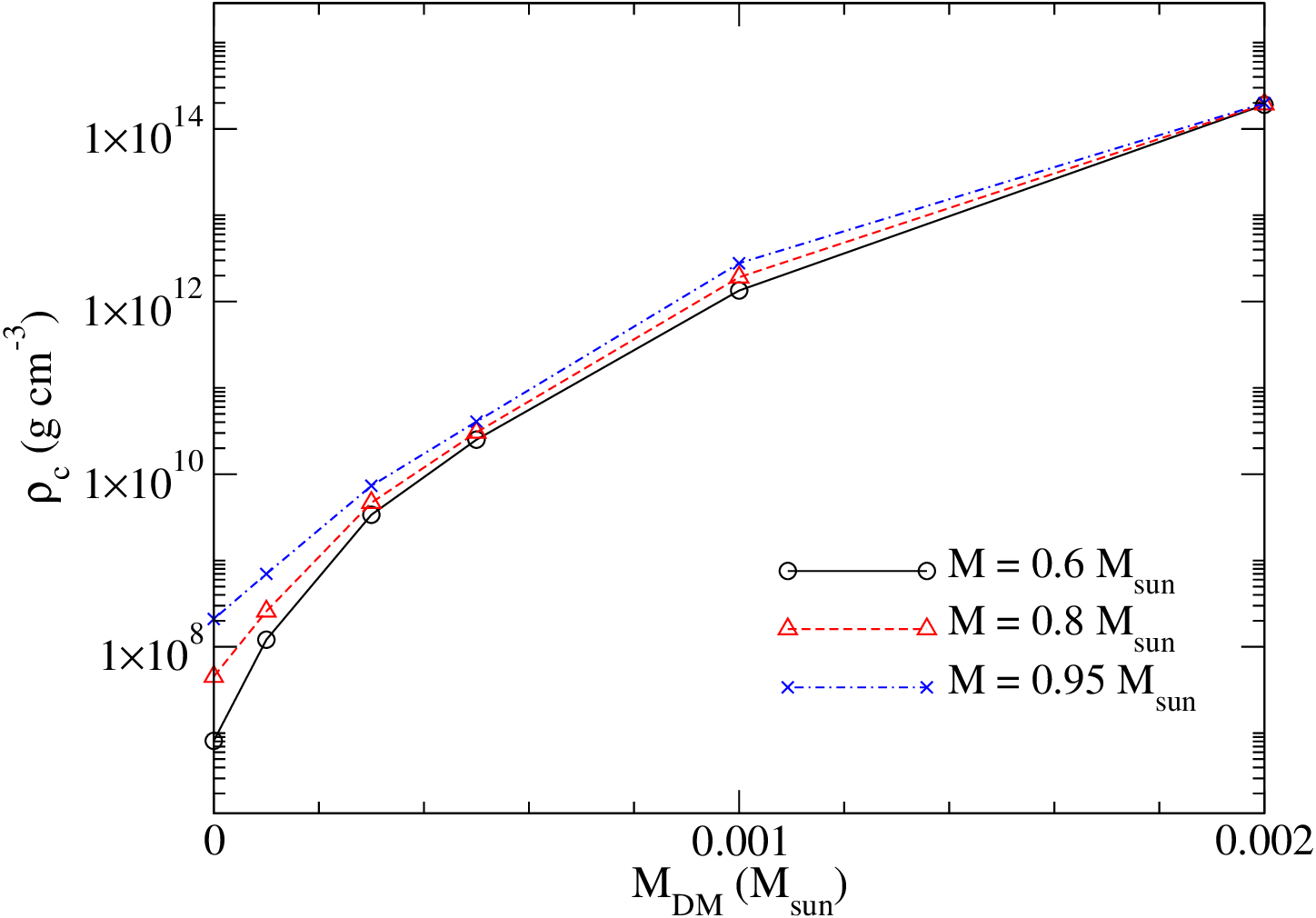}
\caption{NM central density $\rho_c$ is plotted against DM core mass
$M_{{\rm DM}}$ for three sequences of HWD with fixed total mass $M$. The DM particle
mass is fixed at $m_{{\rm DM}}=10$ GeV. }
\label{fig:Mrho_dm10_newplot}
\end{figure}

In Fig.~\ref{fig:Mrhoc_dm10GeV} we plot the total mass $M$ against the NM 
central density $\rho_c$ for $m_{{\rm DM}} = 10$ GeV. 
Three sequences of HWD with different DM core 
masses $M_{{\rm DM}}$ are shown. For comparison, the case without DM (solid line) is 
also plotted. The vertical line at a lower density divides the stable 
(labeled as WD) and unstable branches of traditional WD models, while the 
other vertical line at a higher density marks the onset of the stable branch 
of neutron stars (labeled as NS).  
The region between the two vertical dashed lines is thus the ``forbidden''
region in which no stable traditional WD can exist for our chosen NM EOS model.
It is seen clearly from Fig.~\ref{fig:Mrhoc_dm10GeV} that stable HWD can exist 
in this traditional forbidden region. 
For a given $M$, the NM central density of a HWD increases by a few orders 
of magnitude as $M_{{\rm DM}}$ increases from $10^{-4} M_\odot$ to 
$10^{-3} M_\odot$. In particular, the HWD models with $M_{{\rm DM}}=10^{-3} M_\odot$
have central densities above neutron drip. 
Note, however, that the maximum stable mass of HWD is insensitive to 
$M_{{\rm DM}}$ as we have shown in Fig.~\ref{fig:mr10}. 
To further illustrate how the NM central density changes with $M_{{\rm DM}}$, we 
plot in Fig.~\ref{fig:Mrho_dm10_newplot} $\rho_c$ against $M_{{\rm DM}}$ for three 
different sequences of fixed total mass $M=0.6$, 0.8 and 0.95 $M_\odot$. 
We see that the values of $\rho_c$ for the three sequences get closer as 
$M_{{\rm DM}}$ increases. They can even reach the range of nuclear-matter density 
($\sim 10^{14}\ {\rm g\ cm}^{-3}$) when $M_{{\rm DM}}=2\times 10^{-3} M_\odot$, 
beyond which no stable HWD can be constructed as we have discussed above. 

\begin{figure}
\centering
\includegraphics*[width=8cm,height=7cm]{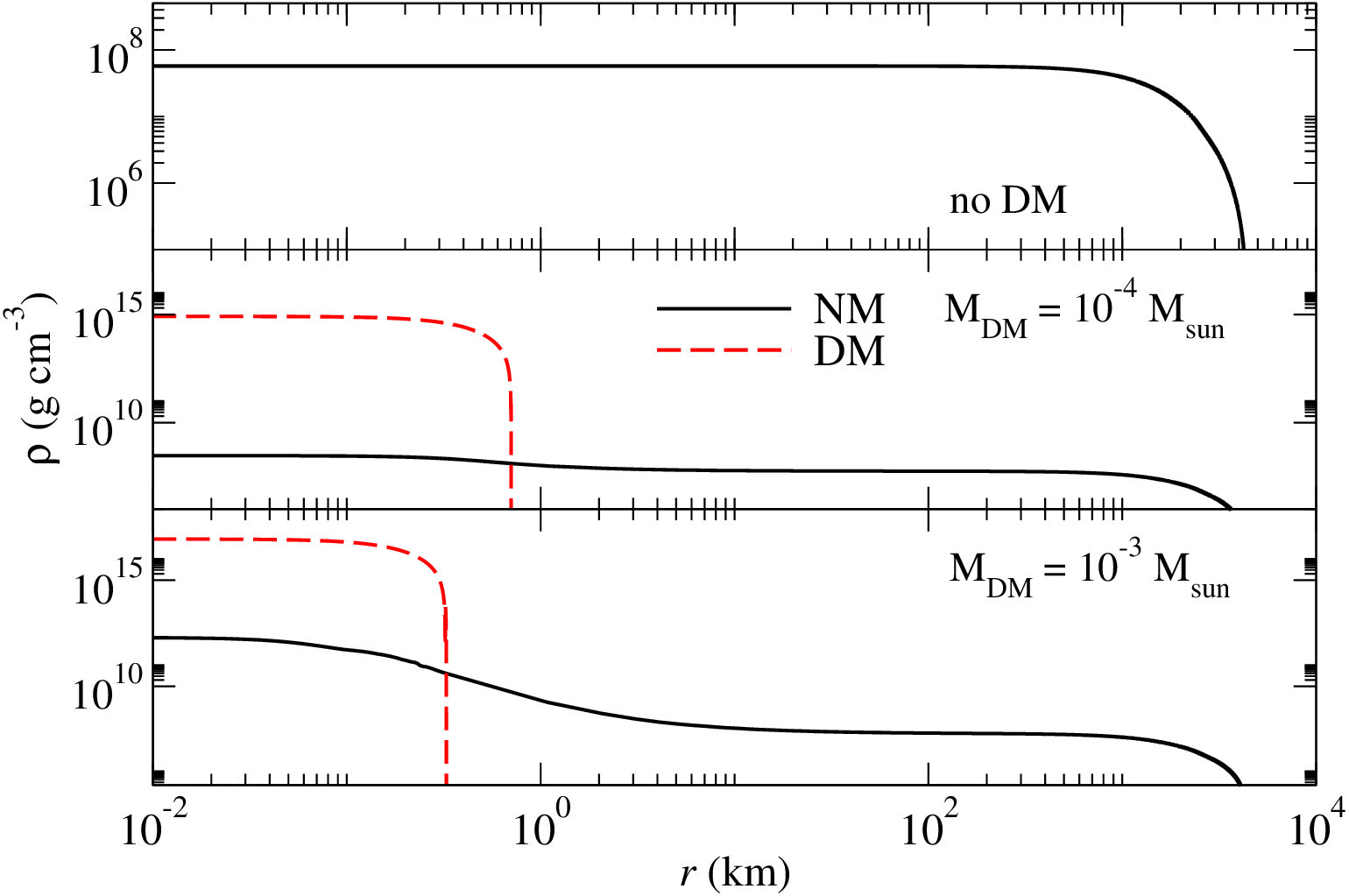}
\caption{ Density profiles for three stellar models with the same 
total mass $M=0.83 M_\odot$. The upper panel shows the profile 
for a traditional WD without DM. The middle and lower panels 
show the profiles of two HWD models with different DM core mass
$M_{{\rm DM}}$. The DM particle mass is fixed at $m_{{\rm DM}} = 10$ GeV.
The solid and dashed lines represent the profiles of NM and DM, 
respectively. 
}
\label{fig:den_profile_nm083}
\end{figure}

In Fig.~\ref{fig:den_profile_nm083}, we plot the density profiles for three 
different stellar models with the same total mass $M=0.83 M_\odot$
and $m_{{\rm DM}} = 10$ GeV.  
The upper panel is the density profile for a traditional WD without DM. 
The middle panel shows the NM (solid line) and DM (dashed line) density 
profiles for a HWD with $M_{{\rm DM}}=10^{-4} M_\odot$, while the bottom panel 
shows the profiles for a HWD with $M_{{\rm DM}}=10^{-3} M_\odot$. 
Fig.~\ref{fig:den_profile_nm083} shows that the presence of a DM core affects 
the distribution of NM near the core significantly. 
As $M_{{\rm DM}}$ increases and the DM core becomes more compact, the NM in the core 
region with a size $\sim 1$ km can be squeezed to density above neutron drip. 
However, slightly outside the DM core, the NM density drops quickly to the 
level commonly found in the core of a traditional WD. It should be noted
that despite the difference in the density profiles in the core regions 
of the HWD and traditional WD models in Fig.~\ref{fig:den_profile_nm083}, 
their global properties such as their masses and radii, are essentially the 
same. 
Our results thus suggest that HWD models with a tiny DM core are consistent 
with many observed WD candidates. However, it also means that it would be  
challenging to distinguish these HWD and traditional WD models 
observationally.

\begin{figure}
\centering
\includegraphics*[width=8cm,height=6cm]{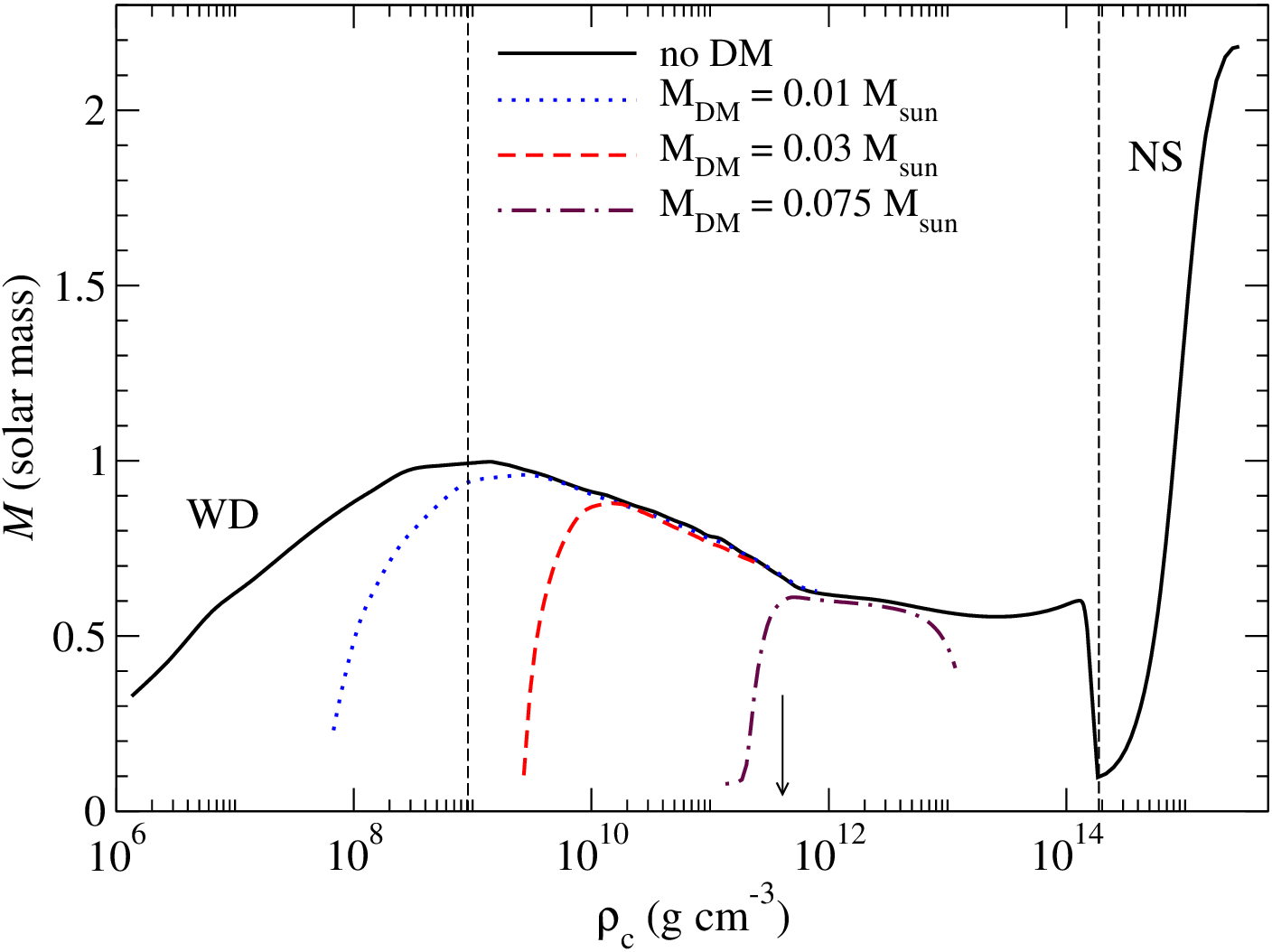}
\caption{Same as Fig.~\ref{fig:Mrhoc_dm10GeV}, but for $m_{{\rm DM}}=1$ GeV. }
\label{fig:Mrhoc_dm1GeV}
\end{figure}

Finally, we plot $M$ against $\rho_c$ for the case $m_{{\rm DM}}=1$ GeV in 
Fig.~\ref{fig:Mrhoc_dm1GeV} for comparison. 
Similar to the case $m_{{\rm DM}}=10$ GeV, we see that the stable branch 
(with $dM/d \rho_c > 0$) of HWD migrates to the 
traditional forbidden region as $M_{{\rm DM}}$ increases. The central density can 
reach above neutron drip for the case $M_{{\rm DM}}=0.075 M_\odot$. 
In contrast to the previous case with $m_{{\rm DM}}=10$ GeV, the maximum stable mass 
of HWD now depends more sensitively on $M_{{\rm DM}}$. 
In particular, it is interesting to note that the curves for HWD models are 
all bound above by the traditional-WD curve. 
We have checked the stability of HWD by analyzing the radial oscillation
modes of these stars as we have done previously for DM admixed neutron stars \cite{DANS2}.
In particular, we show in Appendix A that the DM oscillation modes found in 
\cite{DANS2} also exist in HWD.

\section{Discussion}
\label{sec:discussion}

We have used a general relativistic two-fluid formulation to 
study the equilibrium structure of HWD with DM cores, with the DM particle
mass $m_{DM}$ and the DM core mass $M_{DM}$ as parameters. 
The DM particles are assumed to be non-self-annihilating and form an ideal 
degenerate Fermi gas.

For massive DM particles with $m_{{\rm DM}} = 100$ GeV, we find that stable
HWD can only sustain a tiny DM core with 
$M_{{\rm DM}} \sim 10^{-6} M_\odot$. The masses of these HWD are dominated
by the NM fluid, and hence the global structures of these stars are 
essentially the same as traditional WD without DM. 
If the DM core mass increases beyond $\sim 10^{-6} M_\odot$, the HWD would 
become unstable and collapse promptly to a black hole. 
For less massive DM particles with $m_{{\rm DM}} = 10$ GeV, 
the DM core inside a HWD can be as large as $M_{{\rm DM}} \sim 10^{-3} M_\odot$, 
but still it is not massive enough to affect the global structures, such as 
the mass and radius, of the star significantly. Nevertheless, the 
gravitational attraction of the kilometer-sized DM core can now squeeze the 
NM in the core region to density above neutron drip. In some cases, the NM
central density of the star can even reach the range of nuclear-matter 
density.
The properties of NM in the inner cores of these HWD are very different 
from those of traditional WD which have typical central densities below 
neutron drip. However, from the observational point of view, these HWD and 
traditional WD models could be indistinguishable because their global 
properties are essentially the same. 

In our view, the more interesting result in our 
study is obtained from the case of low-mass DM particles with $m_{{\rm DM}} 
\sim 1$ GeV. The DM core of a HWD in this case 
can reach the level $M_{{\rm DM}} \sim 0.1 M_\odot$ 
and affect the structure 
of the star significantly. In particular, the 
Chandrasekhar mass limit of these HWD depends 
sensitively on $M_{{\rm DM}}$ 
and can decrease by 40\% as we increase $M_{{\rm DM}}$ 
from 0 to $0.075 M_\odot$. Moreover, the radii of
these HWD can be as 
small as $\sim 3000$ km. Comparing to HWD formed 
by massive DM particles, these moon-sized HWD 
could be more easily 
distinguished from traditional WD, which have typical 
radii ranging from about 5000 km to 10000 km. 
Note that while our results presented here 
are based on one particular NM EOS model, we have 
in fact tried different EOS models and found 
that our results still hold qualitatively.

Our study focuses only on non-self-annihilating DM.
In general, other DM particle models may be considered,
such as self-annihilating DM. Also, DM accretion could
be included. However, we remark that  
even if we consider DM accretion, from the analysis in 
\cite{Kouvaris2008}, a compact star
cannot accumulate DM in the mass range of the 
DM core described here in cosmological timescale, 
except when the HWD is embedded in a region of ultra-high DM density. 
For the non-self-annihilating DM we considered in this paper, 
one possible way for a WD to acquire a relatively massive
DM core is that the DM is already trapped inside
the star during its early stage of proto-star formation.
In Appendix B, we provide an order-of-magnitude analysis on 
how much DM can be trapped in that scenario.

In recent years, with the advancement in WD observation using  
double-lined eclipsing binaries, the mass and radius of the WD in 
WD-main-sequence binary systems can now be accurately measured 
\cite{Andersen1991, Southworth2007}. The properties of a number of 
WD have been precisely measured \cite{Parsons2010, Pyrzas2012, Parsons2012a, 
Parsons2012c}. It is thus not inconceivable that moon-sized HWD 
(if exist) could be detected in the near future. 
The detection of such kind of small HWD will be a hint that DM 
particles are low-mass ($m_{{\rm DM}} \sim 1$ GeV) and non-self-annihilating. 
However, we cannot exclude the possibility 
that compact WDs are formed by other mechanisms, 
such as with the help of a strange matter core \cite{Glendenning1994, 
Glendenning1995, Alford2012}.


Finally, let us remark that as the Chandrasekhar mass limit of HWD formed 
by 1-GeV scale DM particles depends sensitively on $M_{DM}$, the initial 
conditions of Type Ia supernovae might not be as universal as generally 
assumed, making it doubtful to what extent Type Ia supernovae 
can be regarded as standard candles - a key assumption in the discovery of 
dark energy.  
It will be interesting to extend our work to investigate how the presence
of a DM core would affect the results of a Type Ia supernova such as its 
luminosity, light curve, and nucleosynthesis yields etc. This will be our 
future investigation.

\begin{appendix}
\section{Radial oscillation Modes of HWD}
\label{sec:Appendix}

In \cite{DANS2} we studied the radial oscillation modes of DM admixed 
neutron stars and found a new class of modes which are characterized 
mainly by the oscillations of DM fluid. 
We have also employed the formulation of \cite{DANS2} to compute the 
oscillation modes of HWD. Similar to the study of DM admixed neutron 
stars, the oscillation modes of HWD can be divided into two classes,
namely the NM fluid modes and the DM fluid modes. The former class 
of modes is driven mainly by NM and depends weakly on the properties 
of DM fluid such as the DM core mass $M_{DM}$. These modes reduce 
properly to the fluid modes of a traditional WD model, 
with the same total mass, as $M_{DM}$ tends to zero.
On the other hand, the second class of modes depend sensitively on the 
DM fluid. 

\begin{figure}
\centering
\includegraphics*[width=8cm,height=6cm]{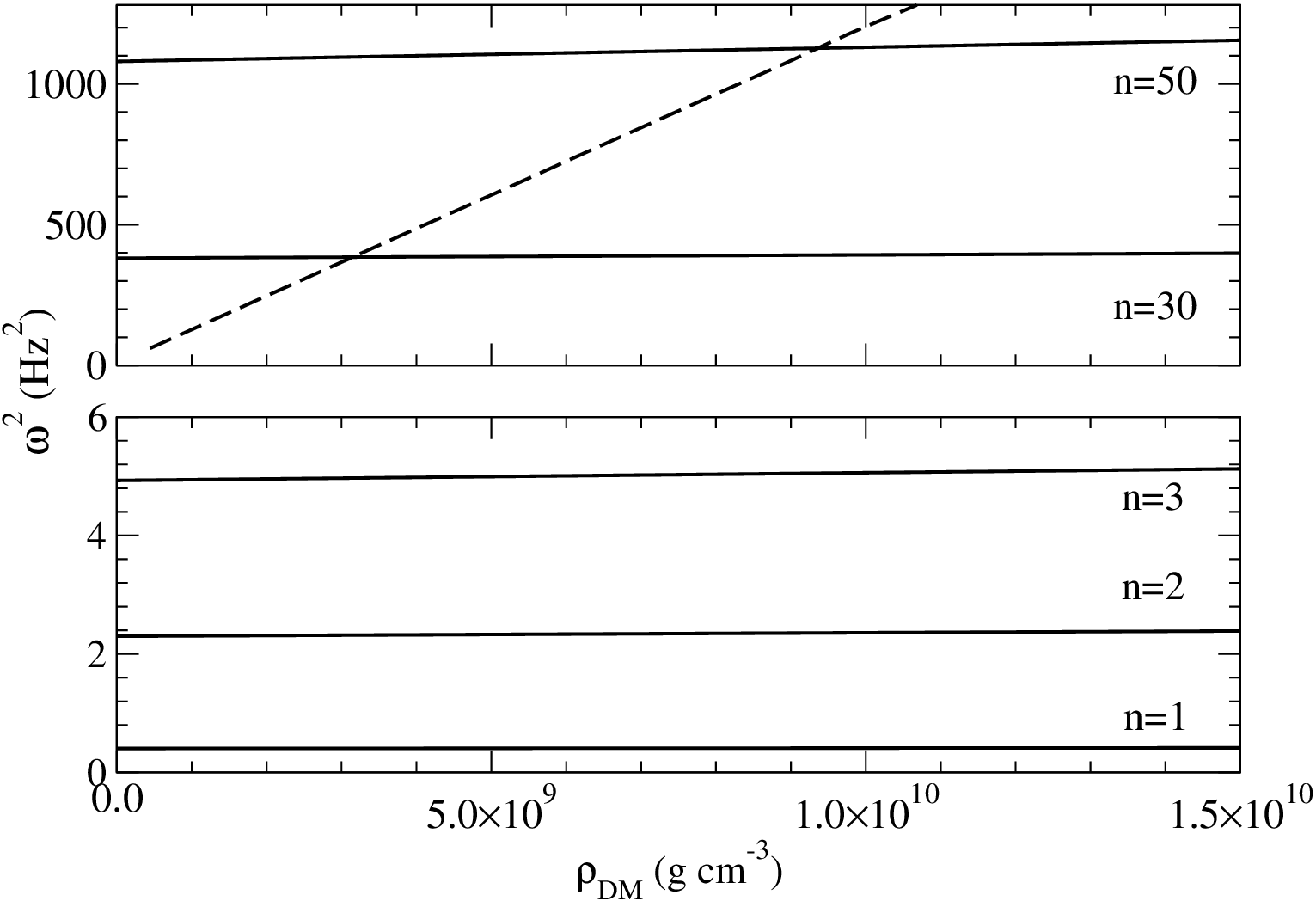}
\caption{The frequency squared $\omega^2$ for different oscillation modes
are plotted against the central DM density $\rho_{{\rm DM}}$. The first three 
NM modes are plotted in the lower panel. In the upper panel, the fundamental 
DM mode (dashed line) and the $n=30$ and $n=50$ NM modes (solid lines) are 
plotted. The total mass of the star is fixed at $M=0.6 M_\odot$ and the DM 
particle mass $m_{{\rm DM}}=1$ GeV.  } 
\label{fig:eigenvalue}
\end{figure}

To illustrate the difference between the two classes of 
modes, we plot in Fig.~\ref{fig:eigenvalue} the mode frequency squared 
$\omega^2$ as a function of the central DM density $\rho_{{\rm DM}}$.
The total mass of the HWD is fixed at $M=0.6 M_\odot$ and the DM particle 
mass is $m_{DM}=1$ GeV. In the 
lower panel of Fig.~\ref{fig:eigenvalue}, we show the first three NM 
modes (solid lines labeled by $n=1$, 2 and 3) and notice that their 
frequencies are of the order of Hz. In the upper panel, we plot the 
fundamental DM fluid (dashed line) and the $n=30$ and $n=50$ NM modes (solid 
lines) for comparison. It is noted that the frequency of the fundamental
DM mode is much higher than that of the NM mode ($n=1$) because the DM 
core is much more compact than the whole star. It can also be seen that the 
DM mode depends much more sensitively on $\rho_{DM}$ as we found previously 
in the case of DM admixed neutron stars \cite{DANS2}.

\section{Possible Formation Mechanisms of HWD}

We notice that there is yet no related study on 
this issue. Here, we make an order-of-magnitude
analysis on possible formation mechanisms of HWD. A stellar object may
acquire DM by accretion \cite{Press1985, Gould1987, Gould1988, Kouvaris2008}. Alternatively,
DM could be trapped in the star formation stage.
We now estimate the amount of trapped DM particles  
in a collapsing molecular cloud of mass $M_{NM0}$,   
with density $\rho_{{\rm NM}}$, temperature $T$ and radius $R$.
We assume that both the densities of NM and DM 
are constant, and the effects of rotation and
magnetic field are neglected. Within the same volume, 
there is also a mass of DM given by
$M_{{\rm DM0}} = \frac{4}{3} \pi R^3 \rho_{{\rm DM}}$.
The total energy of this system is given by
\begin{equation}
E_{{\rm tot}} \sim -\frac{2}{5} \frac{G(M_{{\rm NM0}} + M_{{\rm DM0}})^2}{R} + \frac{3}{2} \frac{M_{{\rm NM0}}}{m_{{\rm H}}} kT.
\label{eq:total_E}
\end{equation}
Here, $m_{{\rm H}}$ is the atomic mass of a hydrogen atom.
Note that we include only the internal energy of NM but not of DM.

Solving Eq. (\ref{eq:total_E}) by requiring $E_{{\rm tot}} = 0$,
we obtain the Jean's radius and then the Jean's mass for
both NM and DM. 
{\it The progenitor of a compact HWD with the mentioned 
$M_{{\rm NM}}$ and $M_{{\rm DM}}$
is a protostar of mass $\sim 1 - 10 M_{\odot}$, with
trapped DM of mass $\sim 10^{-2} M_{\odot}$,
and a mass ratio $M_{{\rm DM0}}/M_{{\rm NM0}} \geq 10^{-2}$. 
This condition can be satisfied for $\rho_{{\rm NM}} \sim 100$ GeV/cm$^{-3}$
and $\rho_{{\rm DM}} > 1$ GeV/cm$^{-3}$, which 
can exist around halo center according to 
observational and N-body simulation results \cite{Ferriere2001, Navarro1996}.}


%

\end{appendix}


\section*{Acknowledgments}
This work is partially supported by a grant from
the Research Grant Council of the Hong Kong
Special Administrative Region, China (Project No. 400910).


\end{document}